\documentclass[
    ,final            % use final for the camera ready runs
  ]
  {aipproc}

\layoutstyle{6x9}
\usepackage{bm}

\newcommand{\ba}{\begin{eqnarray*}}
\newcommand{\ea}{\end{eqnarray*}}

\newcommand{\be}{\begin{equation}}
\newcommand{\ee}{\end{equation}}

\newcommand{\beq}{\[} 
\newcommand{\eeq}{\]}

\newcommand{\la}{\langle}
\newcommand{\ra}{\rangle}
\newcommand{\amp}[1]{\la #1 \ra}

\newcommand{\text}{\hbox}

\newcommand{\simorder}{\raisebox{-4pt}{$\, \stackrel{\textstyle >}{\sim} \,$}}
\newcommand{\simordertwo}{\raisebox{-6pt}{$\, \stackrel{\textstyle <}{\sim} \,$}}

\newcommand{\open}{{<\kern -0.3 em{\scriptscriptstyle )}}}

\begin{document}

\title{\mbox{}\\[-1.8 cm] Gluon saturation effects on single spin asymmetries
\footnote{Presented at the 17th Intl. Spin Physics Symposium,
SPIN2006, Kyoto, Japan, October 2-7, 2006}}

\classification{12.38.-t,12.38.Bx,13.88.+e,13.85.-t\\[-6 mm]}

\author{Dani\"{e}l Boer\\[-2 mm]}{address={Dept.\ of Physics and 
Astronomy, Vrije Universiteit Amsterdam\\
De Boelelaan 1081, 1081 HV Amsterdam,
The Netherlands}
}

\begin{abstract}
We consider forward pion production in $p \, p$ collisions at RHIC energies,
which probes the so-called Extended Geometric Scaling region. Upon inclusion 
of small-$x$ effects via an anomalous dimension within the Color Glass 
Condensate formalism at leading order in $\alpha_s$, a good description of 
the cross section as a function of the transverse momentum of the produced 
pion is obtained. The latter is essential for extractions of the Sivers 
effect from polarized $p \, p$ collisions, since it is a sensitive probe of 
the slope of the cross section. Hence, the presented approach is well suited 
to extract the Sivers effect from single spin asymmetries in 
forward pion production at high energies.
\end{abstract}

\maketitle

%%%%%%%%%%%%%%%%%%%%%%%%%%%%%%%%%%%%%%%%%%%%
%% MAINMATTER
%%%%%%%%%%%%%%%%%%%%%%%%%%%%%%%%%%%%%%%%%%%%

\section{Introduction}
Single transverse-spin asymmetries (SSAs) have been observed in 
$p^{\uparrow} \, p \rightarrow h \, X$
in fixed target experiments \cite{Adams:1991cs} at $\sqrt{s} \approx 20$ GeV 
and in collider experiments at RHIC \cite{Adams:2003fx} 
at $\sqrt{s} = 200$ GeV for hadron rapidities $y_h$ up to 4.
At moderately large $p_t$ of the produced hadron, such SSAs find 
a natural explanation in terms of $\bm{k}_T^{}$-odd transverse momentum
dependent parton distribution functions (TMDs), which 
essentially probe the derivative of the cross
section. Therefore, if there are changes in the cross section, for instance
due to small-$x$ effects, then this may result in changes in the SSAs which
have nothing to do with changes in the spin effect itself. Here the 
small-$x$ effects on SSAs in forward hadron production at RHIC will be 
discussed as an illustration of this point. It is based on Ref.\ 
\cite{Boer:2006rj}. For the spin effect we will restrict to the 
Sivers effect \cite{Sivers:1990cc} (usually denoted by {$\Delta^N
f_{q/p^\uparrow}$} or {$f_{1T}^\perp$}), which is a $\bm{k}_T$-odd TMD. 
As said, it probes the derivative of the cross section which can be seen 
(approximately) as follows: 
\ba 
\lefteqn{A_N \propto d\sigma(p^\uparrow p\rightarrow hX) - d\sigma(p^\downarrow
p\rightarrow hX) \propto 
\int d^2k_t \ {\Delta^N f_{q/p^\uparrow}(x,\vec{k}_t)}\ 
{d\hat\sigma(\vec{q}_t-\vec{k}_t)}}\\[4 mm]
&&\approx {\Delta^N f_{q/p^\uparrow}(x)}
\left[{d\hat\sigma(\vec{q}_t-\amp{k_t}\hat{x})-
d\hat\sigma(\vec{q}_t+\amp{k_t}\hat{x})}\right]
\approx - 2 {\amp{k_t}}\, {\Delta^N f_{q/p^\uparrow}(x)} \,
\hat{x}\cdot \hat{q}_t \,
{\frac{d\hat\sigma(q_t)}{dq_t}},
\ea
where $\hat\sigma=\sigma^{qp\rightarrow q'X}$.  
The first approximation follows from the assumption \cite{Anselmino:1995tv} 
that the Sivers function $\Delta^N f_{q/p^\uparrow}(x,\vec{k}_t)$ is sharply 
peaked around an average transverse momentum (${\amp{k_t} \approx 200}$ {MeV})
that points predominantly in a direction $\hat{x}$ orthogonal to the spin
direction, with a magnitude $\Delta^N f_{q/p^\uparrow}(x)$. The approximations
can also be viewed as resulting from a collinear expansion ($ \amp{k_t} \ll
q_t $).  

The above shows that $A_N$ can increase if the Sivers effect gets
stronger, but also if the average transverse momentum increases or if the 
slope of
the cross section gets steeper. If one does not have a good description of the
cross section, then the extracted Sivers functions from $A_N$ data may not
have the correct magnitude. Therefore, we will first discuss the cross section
description of forward hadron production in $p\,p$ collisions.

\section{Extended geometric scaling region}
If $y_h$ is sufficiently large, then one can probe small $x$ values in
the unpolarized proton and resummation of logarithms in $1/x$ may be 
necessary. In $p \, p \rightarrow h \, X$ at RHIC ($\sqrt{s} = 200$ GeV),  
{$y_h \sim 4$ allows to probe $x \sim 10^{-4}$}. At small $x$ one probes 
mainly gluons and the gluon distribution 
is thought to display saturation (characterized by a scale $Q_s$). For 
$Q_s \simorder 1$ GeV, the Color Glass Condensate (CGC) 
formalism can be employed. HERA data indicate that for $x \sim 10^{-4}$: 
$Q_s \sim 1$ GeV \cite{Golec-Biernat:1998js}.
For ${p}_t \sim Q_s$ saturation effects modify the cross section, 
but even for 
${p}_t$ values significantly above $Q_s$ small-$x$ effects alter the slope
of the cross section w.r.t.\ the standard pQCD treatment. 
For {$Q_s \ \simordertwo \ p_t \ \simordertwo \ Q_{gs} \equiv Q_s^2/\Lambda$} 
--the `extended geometric scaling' region-- quark-CGC scattering is 
well-described by the following replacement in the factorized cross section
description:
\beq 
d\sigma^{qp\rightarrow q'X} \otimes g(x,q_t) \quad \to \quad 
{N_F(x,q_t) \ \propto \ Q_s^2(x) \ 
\text{F.T.} \, (r_t^2)^{{\gamma(x,r_t)}}},
\eeq
where the partonic cross section convoluted with the gluon
distribution is replaced by a dipole forward scattering amplitude $N_F$, which
depends on an anomalous dimension $\gamma$.  
This alters the slope of the cross section w.r.t.\ standard pQCD. 
At large $p_t$, $\gamma$ approaches 
$\gamma_{{\scriptscriptstyle {\rm DGLAP}}}^{} = 1 - {\cal O}(\alpha_s)$.  
The {anomalous dimension $\gamma$} of Refs.\ 
\cite{Dumitru:2005gt,Dumitru:2005kb} follows partly from theory 
and partly from phenomenology, and is given by
\beq
\gamma(x,r_t) = 
\gamma_s + (1-\gamma_s)\, \frac{\log(1/r_t^2Q_s^2(x))} {\lambda
y+d\sqrt{y}+\log(1/r_t^2Q_s^2(x))},
\eeq
with $\gamma_s\simeq 0.627$ (which follows from BFKL evolution with 
saturation boundary conditions), $y=\log 1/x$, $\lambda
\simeq 0.3$ (as obtained from HERA data
\cite{Golec-Biernat:1998js}). The constant 
$d \simeq 1.2$ follows from $d$-$Au$ phenomenology, such that 
the cross section as
function of $p_t$ is well described by the above dipole profile for both
mid and forward rapidity. It describes the slope of the cross section well. 
Overall $p_t$-independent $K$-factors were required, but these do not 
alter the derivative of the cross section and hence are inconsequential for 
our investigation of the SSA. 

\begin{figure}
\begin{minipage}{15cm}
\begin{minipage}{7.2cm}
 \includegraphics[height=.23\textheight]{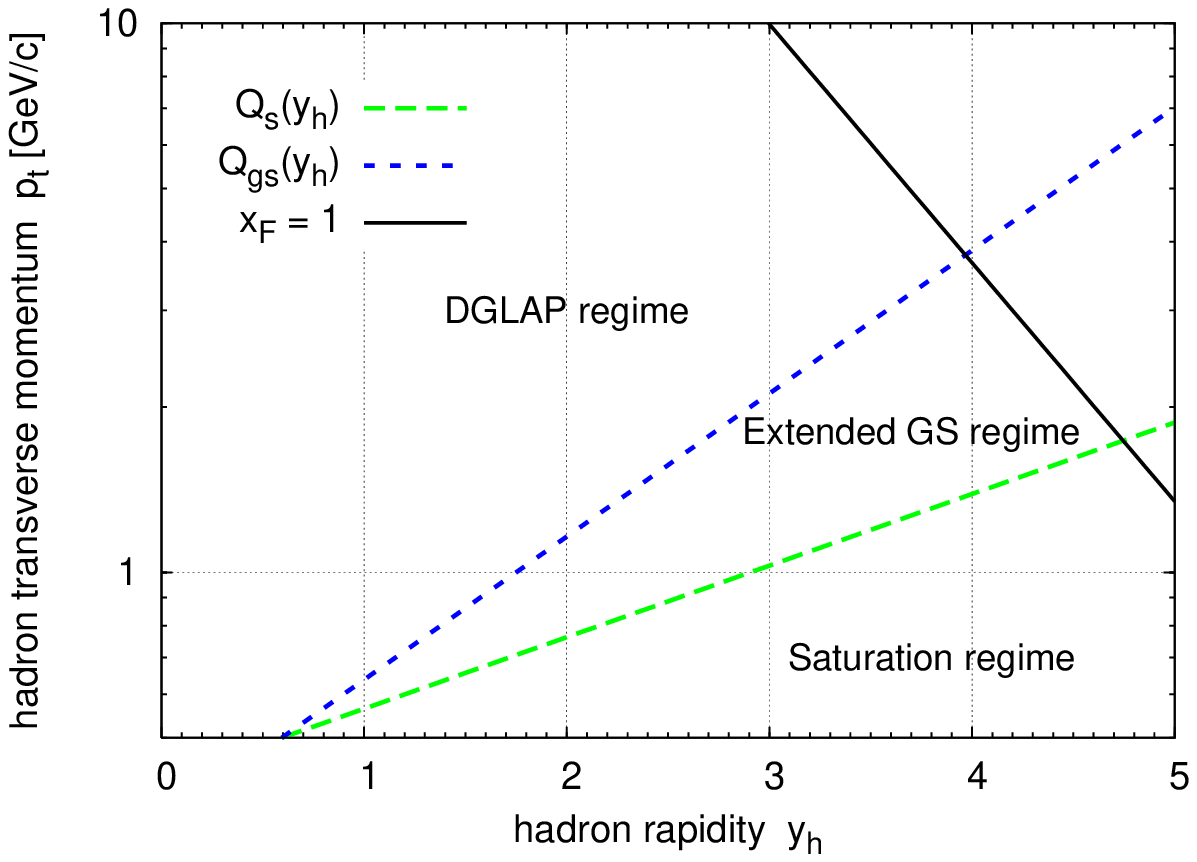}
\end{minipage}
\begin{minipage}{6cm}
 \includegraphics[height=.35\textheight,angle=-90]{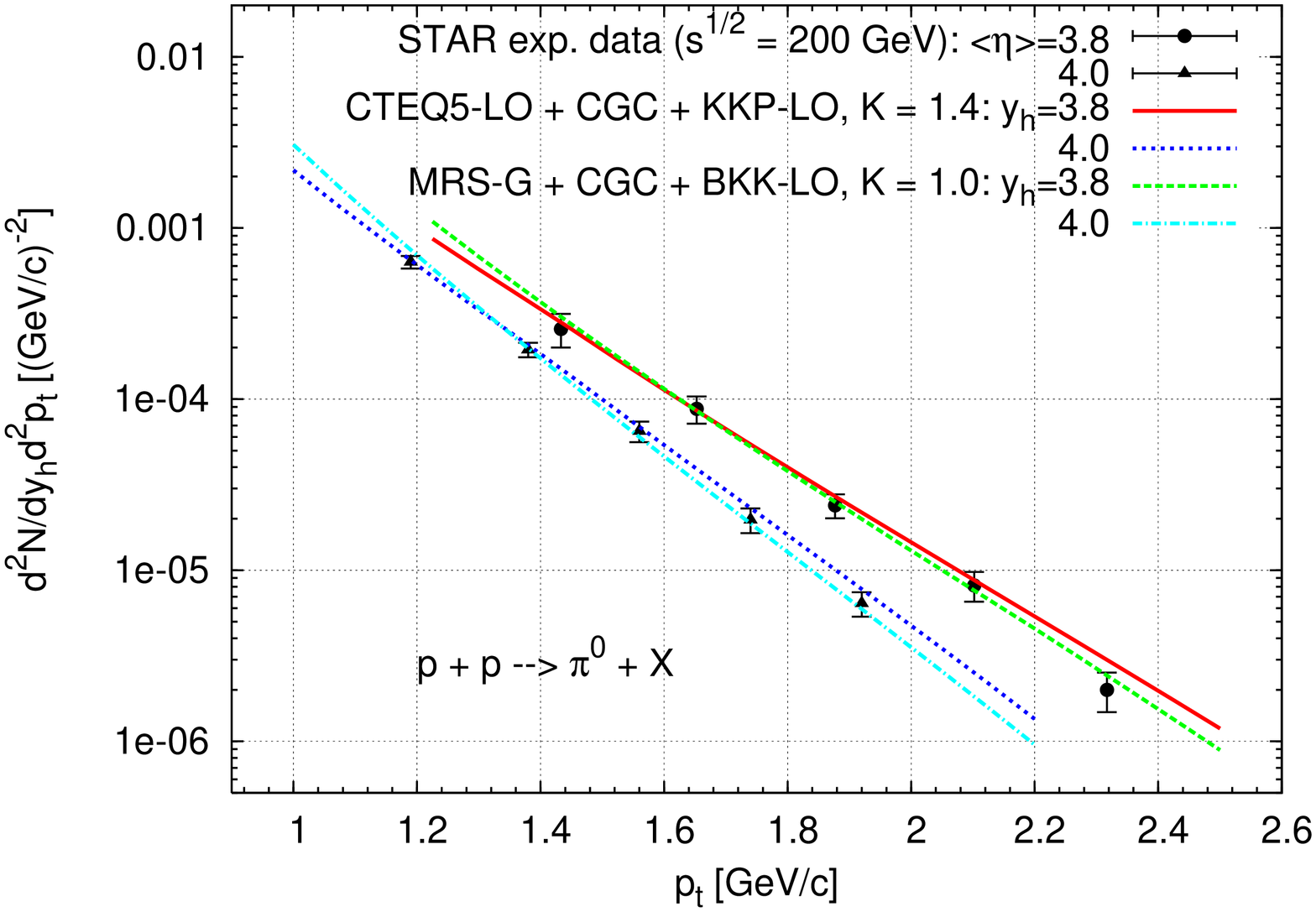}
\end{minipage}
\end{minipage}
 \caption{Left: schematic indication of the extended geometric scaling regime
in the $y_h$-$p_t$ plane for $p\, p$ collisions at RHIC. Right: transverse 
momentum distributions of forward inclusive
$\pi^0$'s 
from unpolarized $p\, p$ collisions at $\sqrt{s}=200$~GeV, compared to
the CGC analysis of Ref.\ \cite{Boer:2006rj}.} 
 \label{phasediagram}
\end{figure}

The question is whether these small-$x$ effects are relevant for
$p \, p$ scattering. It is well-known that NLO pQCD can describe the cross 
section as function of $p_t$ well, except for an indication of a slight 
deviation in slope at very forward rapidities 
\cite{Adams:2006uz,Bland:2006ms}. 
However, the latter cannot be viewed as a
discrepancy given the present uncertainties in the FFs. On the other hand, 
using $Q_s(x) = \left(3 \cdot 10^{-4}/x\right)^{\lambda/2} \, \text{GeV}$ 
from HERA phenomenology and considering 
typical RHIC values for the other parameters, one finds that for 
$y_h\sim 4$, one is in the extended geometric scaling (EGS) region for 
$p_t\ \simorder 1$-$1.5$~GeV/c as shown in Fig.\ \ref{phasediagram} (left
plot). Therefore, the small-$x$ effects are expected to matter.

In Fig.\ \ref{phasediagram} (right plot) one can see
that the CGC formalism in the EGS region can describe the cross
section in that region well, for standard {\em leading order\/} (LO) 
parton distribution and fragmentation functions, even without a 
$K$ factor. Therefore, 
this LO CGC formalism including the anomalous dimension forms a good starting 
point for fits of Sivers functions in this particular kinematic region. 

\section{Single transverse-spin asymmetries}
\begin{figure}
  \includegraphics[height=.4\textheight,angle=-90]{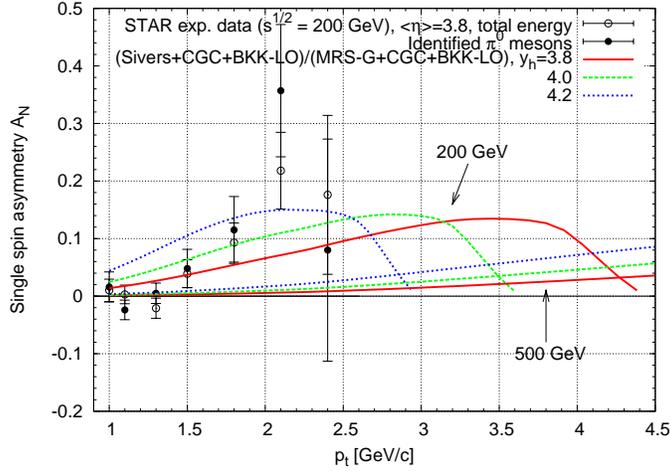}
  \caption{Single transverse-spin asymmetry $A_N$ in the rapidity
interval $y_h=3.8$ - 4.2 for $\sqrt{s}=200$, 500~GeV, with two times 
larger Sivers functions than Ref.~\cite{Anselmino:1998yz}. 
}
\label{SSA}
\end{figure}

From Fig.\ \ref{phasediagram} (right plot) one sees that the slope gets 
steeper as 
$y_h$ increases, such that one expects that the SSA increases with $y_h$ 
accordingly. This is confirmed in Fig.\ \ref{SSA}, where we considered the 
above approach in combination with the Sivers effect. The 
STAR data \cite{Adams:2003fx,Bland:2006ms} 
can be described reasonably by adopting the Sivers function 
parameterization for valence quarks of 
Anselmino \& Murgia \cite{Anselmino:1998yz} {\em times} 2.
Such a quantitative adjustment is not surprising for functions fitted to SSA
data from fixed target experiments, for which the cross section cannot be 
described with pQCD, even at NLO
\cite{Bourrely:2003bw}. The good description of the cross section 
we have obtained indicates that these enhanced Sivers functions may be closer 
in magnitude to the actual Sivers effect. An improved analysis using 
more detailed
transverse momentum dependence for the Sivers functions as investigated by 
Anselmino {\it et al.\/} \cite{Anselmino:2002pd}, seems therefore worth 
doing. 

\section{Conclusions}
At RHIC energies, forward hadron production in $p\, p$ 
scattering is in the extended geometric scaling region, for $y_h \sim 4$ and 
${p}_t \, \simorder \, Q_s \simeq 1$ GeV. 
Here small-$x$ evolution is relevant,
and inclusion of the small-$x$ anomalous dimension is important (even 
essential in this leading order analysis).
The CGC formalism can describe RHIC data (the forward hadron production 
cross section, in particular its derivative) very well. This is important 
for the extraction of Sivers functions
from forward pion SSA, because changes in slope may otherwise be attributed to
changes in the magnitude of the spin effect ($\Delta^N f_{q/p^\uparrow}(x)$) 
or the average transverse momentum. 
We used the STAR data for $p^\uparrow \, p \to \pi^0 \, X$ at 
$\sqrt{s}=200$ GeV and $\amp{\eta}\ (\approx y_h) = 3.8$, to roughly fix the
magnitude of the Sivers functions and studied the $y_h$, $p_t$ and $\sqrt{s}$
dependence of $A_N$ within the outlined approach that incorporates small-$x$
evolution. Details can be found in Ref.\ \cite{Boer:2006rj}. 
 
%%%%%%%%%%%%%%%%%%%%%%%%%%%%%%%%%%%%%%%%%%%%%%%%
%% BACKMATTER
%%%%%%%%%%%%%%%%%%%%%%%%%%%%%%%%%%%%%%%%%%%%%%%%

\begin{theacknowledgments}
The work presented here was done in collaboration with Adrian Dumitru and 
Arata Hayashigaki. This work is part of the research programmes of: 1) the 
`Stichting voor Fundamenteel Onderzoek der Materie (FOM)', which is 
financially supported
by the `Nederlandse Organisatie voor Wetenschappelijk Onderzoek (NWO)'; 2)
the EU Integrated 
Infrastructure Initiative Hadron Physics (RII3-CT-2004-506078). 

\end{theacknowledgments}



\bibliographystyle{aipproc}   % if natbib is available

\bibliography{spin06-proc-hep}

\IfFileExists{\jobname.bbl}{}
 {\typeout{}
  \typeout{******************************************}
  \typeout{** Please run "bibtex \jobname" to optain}
  \typeout{** the bibliography and then re-run LaTeX}
  \typeout{** twice to fix the references!}
  \typeout{******************************************}
  \typeout{}
 }

\end{document}